\documentclass{elsart}
\usepackage{natbib}
\usepackage{amsfonts}
\usepackage{amsmath}
\usepackage{amssymb}
\usepackage{graphicx}
\usepackage{epsfig}
\usepackage{ulem}
\usepackage{color}

\newcommand{\mc}[1]{\mathcal{#1}}
\newcommand{\hmc}[1]{\hat{\mathcal{#1}}}

\def\be{\begin{equation}}
\def\ee{\end{equation}}
\def\bean{\begin{mathletters}\begin{eqnarray}}
\def\eean{\end{eqnarray}\end{mathletters}}
\def\ba{\begin{eqnarray}}
\def\ea{\end{eqnarray}}
\def\bse{\begin{subequations}\begin{align}}
\def\ese{\end{align}\end{subequations}}

\begin{document}
\begin{frontmatter}
\title{Coherent Control of Stationary Light Pulses
\thanksref{X}
}
\author[KL]{F. E. Zimmer}
, \author[Yale]{A. Andr\'{e}}
, \author[Harvard]{M. D. Lukin}
, and \author[KL]{M. Fleischhauer}

\thanks[X]{We dedicate this paper to Bruce W. Shore, one of the fathers 
of the theory of coherent processes in atomic systems, at the occasion of his
70th birthday.}

\address[KL]{Fachbereich Physik, Technische Universit\"at 
Kaiserslautern, 
D 67663 Kaiserslautern, Germany}
\address[Yale]{Physics Department, Yale University, New Haven, 
CT 06520-812, U.S.A.}
\address[Harvard]{Physics Department, Harvard University, Cambridge, 
MA 02138,U.S.A.}
\begin{abstract}
We present a detailed analysis of the recently demonstrated technique 
to generate quasi-stationary pulses of light [M. Bajcsy {\it et al.}, Nature
(London) \textbf{426}, 638 (2003)] 
based on electromagnetically
induced transparency.
We show that the use of counter-propagating control fields to retrieve
a light pulse, previously stored in a collective atomic Raman excitation, 
leads to
quasi-stationary light field that undergoes a slow diffusive spread. 
The underlying physics of this process is identified as pulse matching
of probe and control fields. We then show that
spatially modulated control-field amplitudes allow us to coherently manipulate 
and compress the
spatial shape of the stationary light pulse.
These techniques can provide valuable tools for quantum nonlinear optics and 
quantum information processing.
\end{abstract}


\end{frontmatter}

\section{Introduction}
A very promising avenue towards scalable quantum information systems
is based on photons as information carrier and atomic ensembles as storage
and processing units \cite{Lukin-RMP-2003}. While a number of 
techniques for reliable transfer of quantum information between light
and atomic ensembles have
been proposed \cite{Fleischhauer-PRL-2000,Fleischhauer-PRA-2002,Polzik} 
and in part experimentally realized over the last couple of years
\cite{Polzik,Liu-Nature-2001,Phillips-PRL-2001,vanderWal,Kimble,Kuzmich,Eisaman},
the implementation of quantum information processing in these systems
remains a challenge. This is because deterministic logic operations require
strong nonlinear couplings between few photons or collective excitations 
corresponding to stored photons. To achieve these strong couplings,
long interaction times and tight spatial confinement of the excitations are
needed. Even if long-range interactions between stored photonic qubits are
employed as e.g. in the
dipole-blockade scheme \cite{Lukin-PRL-2001,Cote,Weidemuller} tight confinement
is needed to reach sufficiently high fidelities.

In the present paper we discuss a method that could allow to manipulate the 
spatial shape of a collective excitation corresponding to a stored
light pulse. It is an extension of the recently demonstrated 
technique
to generate quasi-stationary pulses of light 
\cite{Bajcsy-Nature-2003,Andre-PRL-2005} 
in electromagnetically induced transparency (EIT) 
\cite{Fleischhauer-RMP-2005} using 
counter-propagating control fields. In \cite{Bajcsy-Nature-2003} a light pulse
was first stored in a delocalized state of an atomic ensemble by creating 
and adiabatically rotating the collective atom-light excitation, the
so-called dark-state
polariton \cite{Fleischhauer-PRL-2000}, from a freely 
propagating electromagnetic pulse into a stationary 
Zeeman excitation.
The adiabatic rotation, which is accompanied by a decrease of the
group velocity, is facilitated by reducing the intensity of the EIT 
control field. In the form of a pure Zeeman or spin coherence the 
excitation is stored and well protected from the environment 
for rather long times. 
At the same time it is 
however also immobile, thus preventing any manipulation of its spatial shape. 
In addition, the absence of any photonic component to the excitation prevents the use of nonlinear optical interactions for making such excitations interact.
Regenerating a small photonic component of the polariton
by means of a weak stationary retrieval field created by two 
counter-propagating 
lasers, a quasi-stationary pulse of light was created in
the experiment of \cite{Bajcsy-Nature-2003} in a second 
step. Through a 
mechanism known as pulse-matching 
\cite{harris93,Fleischhauer-PRA-1996}, 
the intensity of the regenerated stationary pulse follows the oscillatory
profile of the retrieve laser intensity.
This process allows one to create a stationary excitation with a finite photonic component, i.e., an excitation with stationary, localized electromagnetic energy. One remarkable consequence of this effect is the possibility to enhance nonlinear optical processes \cite{Andre-PRL-2005}.
Another interesting aspect of this technique is that despite the fact that the 
photonic component is at all times very small, the dark-state polariton 
becomes sufficiently mobile to follow the profile of the retrieval 
field. This provides a potential mechanisms to manipulate and control 
the spatial shape of a polariton while keeping most of its 
probability weight in well-protected spin coherences. 

In the present paper a detailed one-dimensional model of the generation of 
quasi-stationary pulses of light by counter-propagating lasers will 
be presented and its predictions compared to 
numerical simulations. It will be shown that in the weak-probe
limit the dynamics of the regenerated light pulse is described by a
set of coupled normal-mode equations \cite{Harris-PRL-1994}
from which exact expressions for the temporal behavior of the 
pulse width can be obtained. It is shown that for sufficiently large
values of the optical depth (OD), these equations reduce to a simple
diffusion equation, with the diffusion coefficient proportional to the
group velocity divided by the optical depth.
It is shown furthermore that control fields with spatially varying 
intensity profiles allow to manipulate the spatial shape of the
stationary light pulse. Using a frequency comb of retrieve fields,
a very narrow stationary mode profile can be generated by a filtering
process. Alternatively using retrieve fields with an intensity difference that 
varies linearly in space in the region of interest
will lead to a stationary field with a Gaussian spatial profile 
and an amplitude exponentially decaying in time. 
A linear dependence of the intensity ratio can be obtained
e.g. by using paraxial 
retrieval beams with spatially displaced foci. Finally we 
demonstrate by numerical examples that moving the laser foci allows to shift
and to compress the stationary pulse of light. Although a quantitative
estimate of the fidelity of such a compression process is not given
here, this shows that stationary pulses of light have a great potential
for the manipulation of the spatial shape of stored photons. 

\section{Stored-light retrieval with counter-propagating control fields}

\subsection{Model}

\begin{figure}[htb]
\begin{center}
\includegraphics[width=9.0cm]{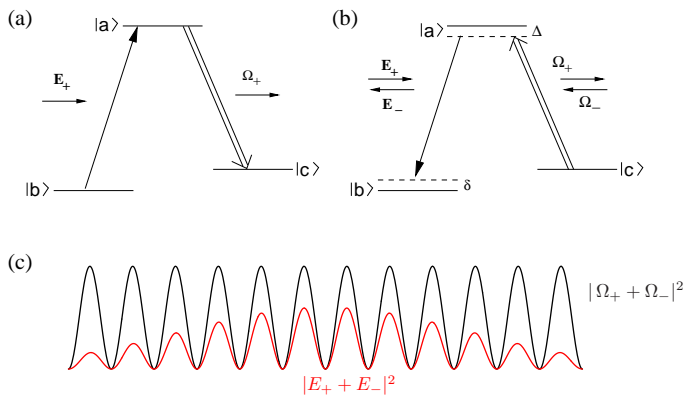}
\caption{Level scheme for storage of light pulse $E_+$ by intensity 
reduction of control field $\Omega_+$ in an EIT system (a), 
and subsequent generation
of a stationary light pulse with two counter-propagating components
$E_\pm$ by applying two counter-propagating control fields $\Omega_\pm$ (b).
The total field distribution of the retrieved light pulse and the control 
fields
is indicated in (c).}
\label{fig-1}
\end{center}
\end{figure}

Let us consider an ensemble of $\Lambda$-type three-level atoms
with one excited level $|a\rangle$ and two lower levels $|b\rangle$ 
and $|c\rangle$. As 
shown in fig.\ref{fig-1}a
the transition 
$|c\rangle-|a\rangle$ of the atoms is coupled to an external drive
field of frequency $\omega_c$ and wavenumber $k_c$ 
characterized by the Rabi-frequency $\Omega_+(z,t)=\Omega_{+0}(z,t)
\exp\{i(k_c z -\omega_c t)\}$. Respectively the transition
$|b\rangle-|a\rangle$ is coupled to a {weak} probe
field of center frequency $\omega$ and wavenumber $k$ 
described by $\textrm{E}_+(z,t)=
E_+(z,t)\exp\{i(k z -\omega t)\}$. The excited state $|a\rangle$
decays radiatively with rate $\gamma$. All other decay and dephasing processes
are neglected. Under conditions of two-photon resonance, i.e. for
$\delta = \omega_{cb}-(\omega-\omega_c)=0$, $\omega_{cb}$ being the
resonance frequency of the lower-level transition, the control field creates
electromagnetically induced transparency (EIT) 
\cite{Fleischhauer-RMP-2005} for the
probe field. Associated with this is a reduction of the group velocity
$v_{\rm gr}$
of a light pulse within a certain frequency range close 
to two-photon resonance:
\begin{equation}
v_{\rm gr}= c\, \cos^2\theta,\qquad \tan^2\theta 
\equiv \frac{g^2 N}{|\Omega_+|^2}. 
\label{eq:group-general}
\end{equation}
Here $g=d/\hbar \sqrt{\hbar\omega_{ab}/2\epsilon_0}$ is the probe-field
coupling strength proportional to the dipole matrix element $d$ of the 
$|a\rangle-|b\rangle$ transition, and $N$ is the density of atoms. An adiabatic
rotation of $\theta=\theta(t)$ from a value close to zero to $\pi/2$ leads
to a slow-down and eventually to a full stop of the probe pulse, which is
associated with a transfer of its quantum state
to a delocalized collective spin (Zeeman) excitation. An adiabatic rotation 
of the mixing angle from $\pi/2$ back to zero 
(or some other value different from $\pi/2$) 
at a later time leads to the retrieval of the stored light pulse 
\cite{Fleischhauer-PRL-2000,Liu-Nature-2001,Phillips-PRL-2001} 
which then propagates in the original
direction with a group velocity determined by eq.(\ref{eq:group-general}).
If for the retrieval a drive beam is used with a different direction, the
stored light is emitted into a direction determined by phase-matching
\cite{Zibrov-PRL-2002,Arve-PRA-2004}.

A very intriguing variant of the retrieval process was suggested and 
experimentally demonstrated in
\cite{Bajcsy-Nature-2003}. Rather than using a single coupling field, two
counter-propagating retrieval beams of the same frequency and intensity 
were used. 
As indicated in fig.\ref{fig-1}b 
this leads to the generation of two counter-propagating probe fields 
$\textrm{E}_\pm(z,t)$
which form a quasi-stationary
standing wave pattern as indicated in fig.\ref{fig-1}c. 
As shown in \cite{Bajcsy-Nature-2003,Andre-PRL-2005} 
the intensity profile of the retrieved field $|E_++E_-|^2$
shows an interference pattern which is similar to that of the control 
field $|\Omega_++\Omega_-|^2$. The initial envelope of this 
pattern is identical to the stored 
light field. Due to a process known
as pulse matching  
\cite{harris93,Fleischhauer-PRA-1996,Harris-PRL-1994}, the 
probe-field envelope
tends to approach that of the control fields with increasing time. 
As a consequence there is a diffusion-like behavior of the retrieved 
field envelope. In the following we want to theoretically analyze
the underlying
physics of this phenomenon within a one-dimensional model. This will then
allow us to discuss a number of interesting generalizations. 
For this we consider the interaction Hamiltonian in a slowly-varying
time frame
\begin{eqnarray}
H &=& -\hbar \int \textrm {d} z\, \biggl[\Delta\sigma_{cc}(z,t) 
+(\delta+\Delta)
\sigma_{aa}(z,t) + \Bigl(\Omega(z,t)\sigma_{ac}(z,t) + h.a.\Bigr),\nonumber\\
&&\qquad\qquad + g\Bigl(E(z,t)\sigma_{ab}(z,t)+h.a.\Bigr)\biggr]
\label{Hamiltonian}
\end{eqnarray}
where $\Delta=\omega_{ac}-\omega_c$ and $\delta=\omega_{cb}-(\omega-\omega_c)$
are the one- and two-photon detuning. The $\sigma_{\mu\nu}$'s are
continuous versions of the slowly-varying (in time) single-atom flip operators
\begin{equation}
\sigma_{\mu\nu}(z,t) = \lim_{\Delta z\to 0}
\frac{1}{\Delta z} \sum_{i\in \Delta z} \sigma_{\mu\nu}^i(t),\qquad 
\sigma_{\mu\nu}^i\equiv |\mu\rangle_{ii}\langle \nu|.
\end{equation}
In the above sum, the atom index $i$ runs over all atoms with positions within
the interval $(z-\Delta z/2,z+\Delta z/2)$. 
$\Omega(z,t)$ and $E(z,t)$ are the (positive-frequency) complex 
Rabi-frequency of the drive field 
and the dimensionless slowly-varying complex 
amplitude of the probe field respectively.
Both can be decomposed into two counter-propagating contributions
\begin{eqnarray}
\Omega(z,t) &=& \Omega_{+0}(z,t)\, {\rm e}^{ik_c z}
+\Omega_{-0}(z,t)\, {\rm e}^{-ik_c z},
\\
E(z,t) &=& E_+(z,t)\, {\rm e}^{ik_c z}+E_-(z,t)\, {\rm e}^{-ik_cz}.
\end{eqnarray}
Note that a fast oscillating term with the retrieve
wavenumber $k_c$ was split off also from the probe field.
Making use of the commutation relations
\begin{equation}
\Bigl[\sigma_{\alpha\beta}(z,t),\sigma_{\mu\nu}(z^\prime,t)\Bigr]
= \delta(z-z^\prime)\, \Bigl(\delta_{\beta\mu}\, \sigma_{\alpha\nu}(z,t)
-\delta_{\alpha\nu}\, \sigma_{\mu\beta}(z,t)\Bigr),
\end{equation}
and considering the perturbative regime of
a weak probe field, where we can set $\sigma_{bb}(z,t)\approx N \approx$ 
const.,
we find the following Langevin equations of motion for the atomic operators:
\begin{eqnarray}
\dot\sigma_{ba} &=& -\bigl(i(\delta+\Delta)+\gamma\bigr)\sigma_{ba}
+i g N \left(E_+\textrm{e}^{ik_cz} + E_-\textrm{e}^{-ik_cz}\right)\nonumber\\
&&
+i\left(\Omega_{+0}\textrm{e}^{ik_c z} +\Omega_{-0}\textrm{e}^{-ik_cz}\right)
\sigma_{bc},\label{Langevin-1}\\
\dot\sigma_{bc} &=& -i\delta\sigma_{bc}
+i\left(\Omega_{+0}^*\textrm{e}^{-ik_c z} +\Omega_{-0}^*\textrm{e}^{ik_cz}
\right)
\sigma_{ba}.
\label{Langevin-2}
\end{eqnarray}
 We have dropped the Langevin noise operator in the equation for 
the optical coherence $\sigma_{ba}$ associated with the
decay rate $\gamma$, since we want to work in the 
adiabatic limit in which this term is negligible \cite{Matsko-PRA-2001}.
The above equations suggest the decomposition of the optical
coherence in two counter-propagating components
$\sigma_{ba}(z)=\sigma_{ba}^{(+)}(z,t)
\textrm{e}^{ik_cz}+\sigma_{ba}^{(-)}(z,t)
\textrm{e}^{-ik_cz}$. Substituting this into (\ref{Langevin-1}) and
(\ref{Langevin-2}) and making a secular approximation, i.e.
collecting terms with the similar oscillatory terms $
(\textrm{e}^{\pm i k_c z})$ and neglecting
fast oscillating contributions $(\textrm{e}^{\pm i 2 k_c z})$, yields
\begin{eqnarray}
\dot\sigma_{ba}^{(\pm)} &=& - \bigl(i(\delta+\Delta)+\gamma\bigr)
\sigma_{ba}^{(\pm)}
+ i g N E_\pm + i\Omega_{\pm 0}
\sigma_{bc},\label{Langevin-11}\\
\dot\sigma_{bc} &=& - i\delta\sigma_{bc}
+ i\Omega_{+0}^*\sigma_{ba}^{(+)}
+ i\Omega_{-0}^*\sigma_{ba}^{(-)}.\label{Langevin-12}
\end{eqnarray}
In the following we assume $\Omega_{\pm 0}^*=\Omega_{\pm 0}$.
Note, that it is also possible to not make the secular approximation,
as is discussed briefly in the appendix.

If the temporal changes of the
slowly varying field amplitudes are slow compared to
$\gamma^{-1}$, we can adiabatically eliminate the optical coherence.
Under these conditions we find
\begin{equation}
\sigma_{ba}^{(\pm)}(z,t) = \frac{{\rm i} g N E_\pm(z,t) +{\rm i} \Omega_{\pm 0}(z,t)
\sigma_{bc}}{i(\Delta+\delta)+\gamma},\label{ad-elimin-opt-coh}
\end{equation}
which leads to the effective equation of motion for the spin coherence
\begin{eqnarray}
\dot\sigma_{bc}=-\left(i\delta +\frac{\Omega_0^2}{i(\Delta+\delta)+\gamma}
\right)\sigma_{bc}-\frac{g N \bigl(E_+\Omega_{+0} 
+E_-\Omega_{-0}\bigr)}{i(\Delta+\delta)+\gamma},\label{ad-elimin-spin}
\end{eqnarray}
where $\Omega_0^2 =\Omega_{+0}^2+\Omega_{-0}^2$.
Eqs.(\ref{ad-elimin-opt-coh},\ref{ad-elimin-spin}) are the 
main equations for the
temporal evolution of the atomic system. 
They describe the dynamics of the 
spin coherence adiabatically followed by the optical coherences.
The second set of equations needed for the description of the
system are the wave equations for the two probe field components
$E_\pm$. In slowly-varying envelope approximation and within the
one-dimensional model considered here, they read
\begin{eqnarray}
\Bigl(\partial_t\pm c\partial_z\Bigr)\, E_\pm(z,t) = -i\Delta\omega \, E_\pm(z,t) + i g \sigma_{ba}^{(\pm)}.\label{Maxwell}
\end{eqnarray}
Here $\Delta\omega=\omega-\omega_c$, and free-space dispersion
$k=\omega/c$ and $k_c=\omega_c/c$
was assumed.

\subsection{effective field equations in the adiabatic limit}

In order to solve the shortened wave equations for the probe fields
coupled to the atomic spins, we now apply an adiabatic perturbation to
the time evolution of the atomic spin coherence, eq.(\ref{ad-elimin-spin}).
In lowest order of this expansion we ignore the time derivative of
the spin coherence $\sigma_{bc}$ altogether. This approximation is however
not sufficient, since it does not capture important effects such as the
group velocity and is only valid if the characteristic pulse
times are large compared to the travel time through the medium.
In order to describe finite group velocities, the first order
correction needs to be taken into account. This yields
\begin{eqnarray}
\sigma_{bc} &\approx & -\frac{gN\bigl(E_+\Omega_{+0} + E_-\Omega_{-0}\bigr)}
{\Omega_0^2+i\delta\Gamma}\nonumber\\
&& +\frac{i(\Delta+\delta)+\gamma}{\left(\Omega_0^2
+i\delta\Gamma\right)^2} gN
\bigl(\Omega_{+0}\partial_t E_+ + \Omega_{-0}\partial_t E_-\bigr),
\label{sigma-bc}
\end{eqnarray}
where we have disregarded time derivatives of $\Omega_{\pm 0}$ and have 
introduced the complex parameter $\Gamma\equiv \gamma+i(\Delta+\delta)$. 
Substituting this result into the expression for the 
optical coherence $\sigma_{ba}$ and
subsequently into the wave equations (\ref{Maxwell}) eventually leads to the 
coupled field equations
\begin{eqnarray}
\Bigl(\partial_t\pm c\partial_z\Bigr) \, E_\pm &=& -\left(i\Delta\omega
+\frac{g^2 N}{\Gamma}\right)\, E_\pm
+\frac{g^2 N \Omega_{\pm 0}\bigl(\Omega_{+0}\, E_+
+\Omega_{-0}\, E_-\bigr)}{\Gamma\bigl(\Omega_0^2+i\delta\Gamma\bigr)}
\nonumber\\
&& -\frac{g^2 N\Omega_{\pm 0}\bigl(\Omega_{+0}\, \partial_tE_+
+\Omega_{-0}\, \partial_t E_-\bigr)}{\bigl(\Omega_0^2+i\delta\Gamma\bigr)^2}.
\end{eqnarray}
These equations can be written in a more transparent form by introducing the
mixing angles $\theta$ and $\phi$
\begin{equation}
\tan^2\theta \equiv \frac{g^2 N}{\Omega_0^2},\qquad
\tan^2\phi\equiv \frac{|\Omega_{-0}|^2}{|\Omega_{+0}|^2},
\label{Def-angles}
\end{equation}
and by assuming a small two-photon detuning, i.e.
$\Omega_0^2\gg |\delta\Gamma|$:
\begin{eqnarray}
&&\Bigl(\partial_t + c\cos^2\theta\partial_z\Bigr)E_+
= -i\bigl(\Delta\omega \cos^2\theta+\delta \sin^2\theta\bigr)E_+
\nonumber\\
&&\qquad\qquad +
\sin^2\theta\sin\phi\Bigl(\sin\phi\partial_t E_+ -\cos\phi\partial_t E_-
\Bigr) \label{Maxwell-E+}\\
&&\qquad\qquad
- \cos^2\theta \frac{g^2 N}{\Gamma} \sin\phi\bigl(\sin\phi E_+-\cos\phi
E_-\bigr),\nonumber
\end{eqnarray}
and
\begin{eqnarray}
&&\Bigl(\partial_t - c\cos^2\theta\partial_z\Bigr)E_-
=  -i\bigl(\Delta\omega \cos^2\theta+\delta \sin^2\theta\bigr)E_-
\nonumber\\
&&\qquad\qquad
-\sin^2\theta\cos\phi\Bigl(\sin\phi\partial_t E_+ -\cos\phi\partial_t E_-
\Bigr) \label{Maxwell-E-}\\
&&\qquad\qquad
+ \cos^2\theta \frac{g^2 N}{\Gamma} 
\cos\phi\bigl(\sin\phi E_+-\cos\phi E_-\bigr).\nonumber
\end{eqnarray}
One recognizes that the two field components propagate with an effective group
velocity $v_{\rm gr}=c\, \cos^2\theta$ similar to eq.(\ref{eq:group-general}).
The first bracket on the right hand side of eqs.(\ref{Maxwell-E+}) and
(\ref{Maxwell-E-}) represents a phase-mismatch, which however vanishes
for a two-photon detuning chosen such that 
\begin{equation}
\delta=-\Delta\omega \cot^2\theta\approx -\Delta\omega\frac{v_{\rm gr}}{c}.
\label{phase-matching}
\end{equation}
If  $|v_{\rm gr}|\ll c$
the two-photon detuning is very small and does not lead to a violation of the
EIT condition. In the retrieval process the probe field will build up
with a center frequency such that the phase-matching condition 
(\ref{phase-matching}) is fulfilled.

\subsection{normal modes and pulse matching}

The structure of the two field equations (\ref{Maxwell-E+}) and
(\ref{Maxwell-E-}) suggests the introduction of
the two normal modes \cite{Harris-PRL-1994}
\begin{eqnarray}
E_S \equiv \cos\phi\, E_+ +\sin\phi\, E_-,\qquad
E_D \equiv \sin\phi\, E_+ -\cos\phi\, E_-,\label{normal-modes}
\end{eqnarray}
which we denote as sum and difference normal mode.
In terms of these modes the propagation equations read
\begin{eqnarray}
&&\Bigl(\partial_t+v_{\rm gr}\cos(2\phi)\partial_z\Bigr)\, E_S + 
v_{\rm gr}\sin(2\phi)\partial_z\, E_D=
\nonumber\\
&&\qquad\qquad \qquad +v_{\rm gr}\bigl(\partial_z\phi\bigr)\bigl(\sin(2\phi) E_S -
\cos(2\phi) E_D\bigr),\label{Maxwell-E-S}\\
&&\Bigl(\partial_t-c\cos(2\phi)\partial_z\Bigr)\, E_D +
c\sin(2\phi)\partial_z\, E_S= 
-\frac{g^2 N}{\Gamma} E_D\nonumber\\
&&\qquad\qquad\qquad -c\bigl(\partial_z\phi\bigr)\bigl(\cos(2\phi) E_S +
\sin(2\phi) E_D\bigr).\label{Maxwell-E-D}
\end{eqnarray}
Here we have assumed that the mixing angles $\theta$ and $\phi$ can be
space dependent but are constant in time.
At the same time, in lowest order of the adiabatic expansion,
 the atomic spin coherence follows the evolution
of the sum normal mode $E_S$, hence we find from eq.(\ref{sigma-bc})
\begin{equation}
\sigma_{bc}=-\sqrt{N}\tan\theta\, E_S.
\label{SigmaBCAdiabaticLimit}
\end{equation}
One recognizes from (\ref{Maxwell-E-S}) and (\ref{Maxwell-E-D})
that apart from the coupling between the normal modes
$E_S$ and $E_D$, the difference mode $E_D$ is strongly absorbed due to the term
$g^2 N/\Gamma$ on the right hand side. As a 
consequence the amplitudes of the retrieved fields approach a 
configuration where $E_D\rightarrow 0$, i.e. a 
configuration where the probe
amplitudes match those of the drive fields:
\begin{equation}
\frac{E_+}{E_-}\rightarrow \cot\phi=\frac{\Omega_+}{\Omega_-}.
\end{equation}
This phenomenon called pulse-matching is well-known for EIT systems
\cite{harris93,Harris-PRL-1994,Fleischhauer-PRA-1996}. 

One recognizes that for almost identical control fields, 
i.~e.~$\Omega_+\approx\Omega_-$ the group velocity 
of the sum mode $E_S$ and difference mode $E_D$ are very small 
due to the $\cos(2\phi)$-term. 
It is zero if 
the strength of the two counter-propagating control beams is
exactly the same. 
If $\Omega_+$ is bigger than $\Omega_-$ the sum mode moves to
the $+z$ direction and vice versa. 
Finally even if the ratio of the control field
envelopes is spatially constant, i.e. if $\partial_z\phi=0$, a small
amplitude of the difference mode will be generated out of the sum mode
due to the term proportional to $\partial_z E_S$
until $E_S$ is constant in space. This coupling will give rise to
the slow spatio-temporal evolution discussed 
in the following sections.

\section{Quasi-stationary pulses of light from spatially homogeneous
retrieval beams}

Let us first consider the case of two spatially homogeneous 
control fields with equal intensity, i.e.
\begin{equation}
\cos(2\phi)=0,\qquad \partial_z\phi=0.
\end{equation}
This can be realized e.g. by two laser beams of the same 
intensity with a negligible curvature of the phase fronts, 
i.~e. in the plain wave regime.
In this case the propagation equations for the sum and difference
mode (\ref{Maxwell-E-S}) and (\ref{Maxwell-E-D}) simplify to
\begin{eqnarray}
\partial_t E_S &=& -v_{\rm gr} \partial_z E_D,\label{Maxwell-ES-homog}\\
\partial_t E_D &=& -c \partial_z E_S -\frac{g^2 N}{\Gamma} E_D.
\label{Maxwell-ED-homog}
\end{eqnarray}
%
%

\subsection{Adiabatic elimination of difference normal mode and 
diffusion equation for resonant probe fields}

\begin{figure}[htb]
\begin{center}
\includegraphics[width=12.0cm]{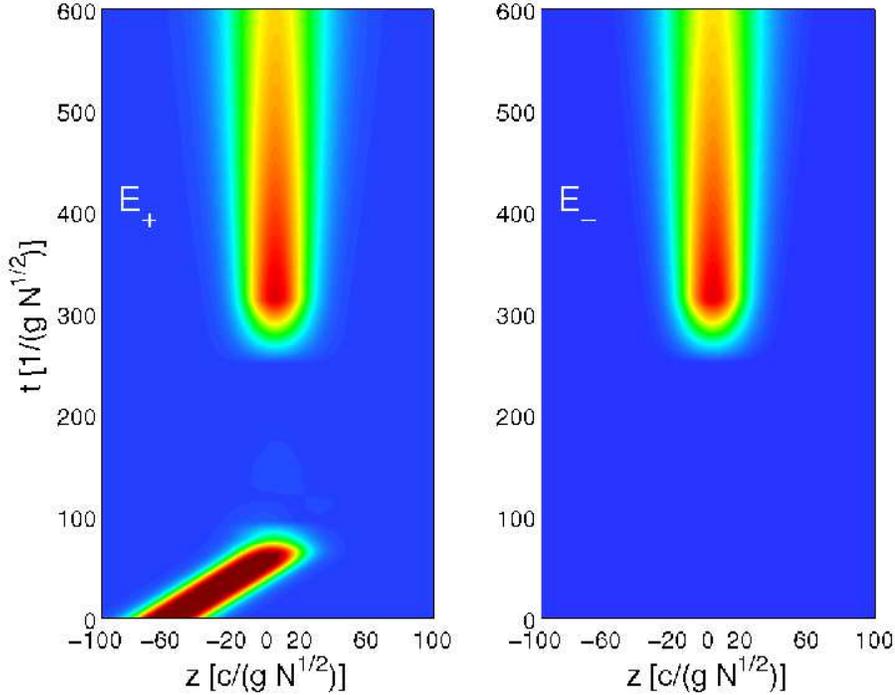}
\caption{Numerical simulation of storage and partial retrieval of a light
pulse, where in the retrieval process two counter-propagating drive beams
with equal and spatially homogeneous intensities have been applied.
For the storage and retrieval we have used 
$\cos^2\theta_+=  0.5*[1-\tanh(0.1*(t-65))]
+0.5*1/3*[1+\tanh(0.1*(t-300))]$, where $\tan^2\theta_\pm=g^2 N/
\Omega_{\pm 0}^2$. For $\cos^2\theta_-$ only the second term
was used.
The color code represents the amplitude of the forward ($E_+$) and 
backward ($E_-$) propagating components of the probe field.
The parameters used for the numerical simulation are: 
$\delta=\Delta=\Delta\omega=0$ and $\gamma=1$ and the 
initial width of the Gaussian wave-packet was $\Delta z(t=0)=10$.}
\label{fig-2}
\end{center}
\end{figure}

Let us consider the case where the drive field detuning $\Delta$ 
is chosen such that the probe fields are resonant, i.e. $\Delta+\delta=0$.
Since for an optically dense medium the phase matching condition
(\ref{phase-matching}) requires only a very small two-photon detuning, this 
is essentially equivalent to the case of resonant drive fields.
Then $\Gamma=\gamma +i(\Delta+\delta)=\gamma$ and
eq.(\ref{Maxwell-ED-homog}) shows that the difference normal mode 
is damped with a rate $g^2 N/\gamma
=c/l_{\rm abs}$, where $l_{\rm abs}$ is the absorption length of the medium
in the absence of EIT. For a medium with sufficiently high optical density
$OD=l/l_{\rm abs}$
the absorption length is typically on the mm scale and thus the
decay time is on the order of a few picoseconds. The typical pulse 
times in light storage
experiments are however much larger. Thus an adiabatic elimination
of the difference normal mode,
i.e. neglecting $\partial_t E_D$ as compared to $(g^2 N/\gamma)E_D$, seems 
justified. Such an elimination leads to
\begin{equation}
E_D=- l_{\rm abs} \frac{\partial}{\partial z}E_S.
\label{EDAdiabaticLimit}
\end{equation}
Making use of this approximation
we find  for the sum normal mode $E_S$ a simple diffusion equation
\begin{equation}
\frac{\partial}{\partial t} E_S = v_{\rm gr} l_{\rm abs}\, \frac{\partial^2}{\partial z^2} E_S = D\, 
\frac{\partial^2}{\partial z^2} E_S
.\label{diffusion}
\end{equation}
In the retrieval process two counter-propagating probe field
components are created with an initial envelope given by the
stored spin excitation. These components then undergo a diffusion process
with a diffusion constant $D=v_{\rm gr} l_{\rm abs}$ given by the product 
of group velocity and absorption length. 
This is illustrated in fig.\ref{fig-2} where false-color images show the
two field distributions $E_+(z,t)$ and $E_-(z,t)$ for a storage process
followed by a partial retrieve with two homogeneous, counter-propagating drive
beams of equal amplitude. The data are obtained from a 
numerical solution of
the wave equation
(\ref{Maxwell}) as well as the full set of atomic density matrix equations 
in secular approximation. The predicted diffusive behavior is 
nicely reproduced.

In the diffusion process the width of the probe field
distribution as well as that of the collective spin excitation
(see eq.(\ref{SigmaBCAdiabaticLimit})) increase according to
\begin{equation}
\Delta z^2(t) =\Delta z^2(t_0) + 2 D (t-t_0).
\end{equation}
Associated with this is a decrease of the excitation density.
Since in a diffusion process the spatial integral of the field is
constant but not the integral of the square of the field, representing
the number of photons,
there is also a (non-exponential) decay of the total number 
of excitations.
After the control fields are switched on again,
the sum mode has a Gaussian shape with width $\Delta z(0)$, and the total
excitation, i.e.
in the retrieved fields and the collective spin, evolves according to
\begin{equation}
n_{\rm tot}(t)= n_{\rm tot}(0) 
\frac{\Delta z(0)}{\sqrt{\Delta z^2(0) + 2 D t}}.
\end{equation}
Thus in order to have negligible losses, the time over which 
a stationary pulse can be maintained is limited by
\begin{equation}
t\ll \frac{\Delta z^2(0)}{D}=\frac{\Delta z^2(0)}{v_{\rm gr}l_{\rm abs}},
\end{equation}
which is exactly the characteristic time for the spread of the
initial wave-packet.

\subsection{Small optical depth}

\begin{figure}[htb]
\begin{center}
\includegraphics[width=9.0cm]{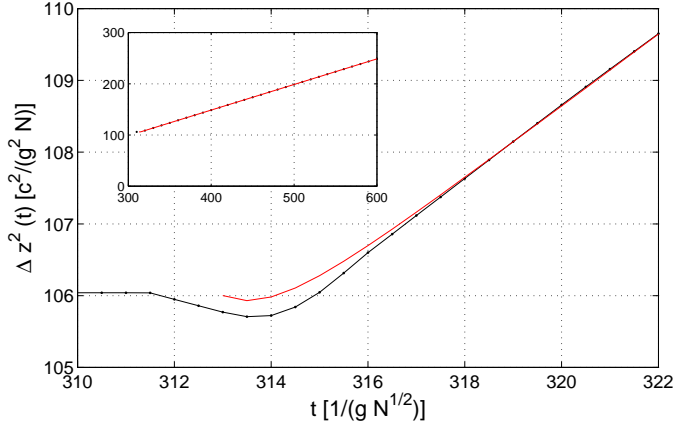}
\caption{Comparison between analytic prediction for
$\Delta z^2(t)$
of the stationary wave-packet of light from eq.(\ref{delta-solution}) with
a numerical simulation of the coupled Maxwell-Bloch equations.
The data are based on the simulation of fig.\ref{fig-2}.
The insert shows the behavior on a larger time scale.}
\label{fig-3}
\end{center}
\end{figure}

If the optical depth of the medium is small,
the adiabatic elimination of $E_D$ may be no
longer justified. It is still possible however to find analytic results
for the moments of the stationary light field.
First of all one finds from eqs.(\ref{Maxwell-ES-homog}) and 
(\ref{Maxwell-ED-homog}) that like in the diffusion limit discussed above,
the integral of $E_S$ is a constant of motion since
$E_D(z=\pm\infty)=0$:
\begin{equation}
\frac{{\rm d}}{{\rm d}t}A(t)=\frac{{\rm d}}{{\rm d}t}
\int_{-\infty}^\infty\!\!{\rm d}z\, E_S(z,t)=0.
\end{equation}
Assuming an initially symmetric spin excitation around $z=0$, the width
of the retrieved light beam is given by the second moment of the sum mode 
$d(t)\equiv\Delta z^2(t)= \int_{-\infty}^\infty 
\,{\rm d}z\, z^2\, E_S(z,t)/A$, which
is coupled to the first moment of the difference mode
$g(t)\equiv \int_{-\infty}^\infty\, {\rm d}z\, z\, E_D(z,t)/A$:
\begin{eqnarray}
\frac{{\rm d}}{{\rm d}t} d(t) &=& 2 v_{\rm gr}g(t),\\
\frac{{\rm d}}{{\rm d}t} g(t) &=& -\frac{c}{l_{\rm abs}} g(t)
+c .
\end{eqnarray}
The solution of these equations can easily be found and reads
\begin{equation}
d(t)=d(0) + 2 D t
+2 D\frac{l_{\rm abs}}{c} \left(1-\frac{g(0)}{l_{\rm abs}}\right)
\left({\rm e}^{-c t/l_{\rm abs}}-1\right).
\label{delta-solution}
\end{equation}
One recognizes that a small absorption length $l_{\rm abs}$ only
affects the short-time evolution.
In fig.\ref{fig-3} a comparison between the analytical prediction 
for $\delta(t)$ obtained from
eq.(\ref{delta-solution}) with $v_{\rm gr}\to v_{\rm gr}(t)$ 
and a numerical simulation of the Maxwell-Bloch equations
is shown. Apart from a short initial time period, where due
to the time dependence of $v_{\rm gr}(t)$ non-adiabatic
couplings lead to small deviations, there is a nearly perfect
agreement between analytic prediction and numerical simulation.

\subsection{Non-equal but constant drive intensities}

If the intensities of the counter-propagating retrieve beams are not equal
but constant in time and space, one has
\begin{equation}
\cos(2\phi)\ne 0,\qquad \partial_z\phi=0.
\end{equation}
As a consequence the equation of motion of the sum normal mode attains,
after adiabatic elimination of the difference mode, a finite drift term
\begin{equation}
\Bigl(\partial_t +v_{\rm gr}\cos(2\phi)\partial_z\Bigr)\, E_S = 
v_{\rm gr}l_{\rm abs}\sin^2(2\phi)\, \partial_z^2 E_S.
\end{equation}
Transforming into a moving frame with $z^\prime = z -v_{\rm gr}\cos(2\phi) t$
and $t^\prime=t$ leads again to a diffusion equation with 
a modified diffusion constant 
$\tilde{D}=v_{\rm gr}l_{\rm abs}\sin^2(2\phi)$. 
Thus in the case of non equal, but constant 
drive intensities,
the diffusive behavior of the quasi-stationary light is superimposed by
a drift motion with a small 
velocity $v_{\rm gr}\cos(2\phi)$. This can be understood in
a very intuitive way. If one of the drive fields is stronger than the other
one, Raman scattering occurs with higher probability into the probe mode
co-propagating with the stronger drive field which causes a drift motion of
the quasi-stationary wave-packet.

\section{Stationary pulses of light generated by spatially 
modulated retrieve fields}

In this section, we discuss two techniques to manipulate the shape of 
stationary pulses with the ultimate goal of confining the pulse to very 
short spatial dimensions. Initially the pulse is stored as a spin coherence 
with a spatial envelope that extends over many wavelengths. We have seen in the
previous sections that the (partial) retrieval of the stored
light pulse by counter-propagating control fields leads to quasi-stationary
pulses of light. The shape of these stationary pulses is determined by 
the envelope of the initial spin coherence as well as the control-field
envelopes through the mechanism of pulse-matching. 

\begin{figure}[htb]    
\begin{center}
\includegraphics[width=8 cm]{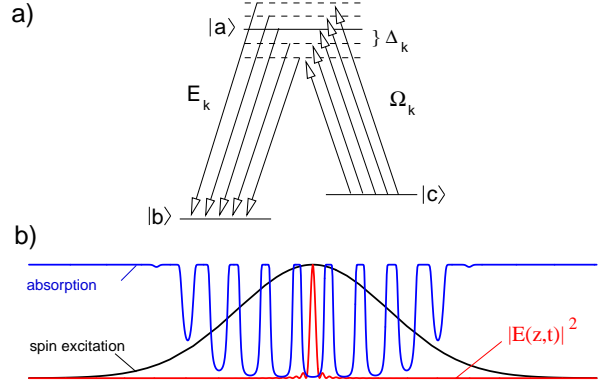}
\end{center}
\caption{\label{fig:shapefig1}
(a) Atomic level structure with multiple-frequency-component 
control field $\Omega(z,t)=\sum_{k}\Omega_ke^{-i\Delta_k(t-z/c)}$, 
and generated signal field $E(z,t)=\sum_k {E}_k(z,t)e^{-i\Delta_k(t-z/c)}$.
(b) Generated signal field envelope, 
showing tight localization due to constructively interfering
frequency components (frequency comb).
}
\end{figure}

This suggests two different mechanisms to manipulate the shape of the
stationary pulse of light. In the first method the atoms 
are illuminated by a frequency comb, i.e.
with control beams that have multiple frequency components of equal
intensity in the forward and backward direction. 
In this way many corresponding frequency components 
are generated for the signal field. These components interfere to create a 
very sharp spatial envelope, which is matched to the sharp
spatial envelope of the control field, potentially 
confined over only a few wavelengths 
(see Fig.~\ref{fig:shapefig1}).

\begin{figure}[htb]    
\begin{center}
\includegraphics[width=8 cm]{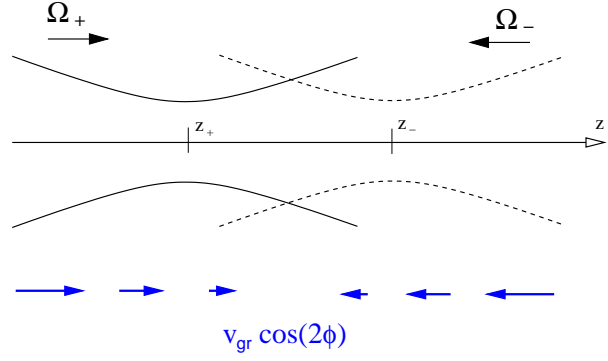}
\end{center}
\caption{\label{fig:cavity}
Paraxial retrieve lasers with spatially varying, Gaussian intensity
profiles determined by the position of the 
focal points $z_+\ne z_-$ create
a spatially varying effective group velocity.
}
\end{figure}

A second method employs a spatial modulation in the difference
of the forward and backward retrieve intensities.
In the last section we have seen that unequal retrieve intensities
can lead to a drift motion of the stationary field with an effective
group velocity $v_{\rm gr}\cos(2\phi)$. If e.g. $v_{\rm gr}\cos(2\phi)$
would be negative for positive
values of $z$ and positive for negative values of $z$, the associated
drift motion would tend to spatially compress the stationary field. 
As will be shown
this can compensate the diffusive spread found in
the last section. This situation, indicated in fig.~\ref{fig:cavity},
 can be achieved when $\phi$ (and in general
also $\theta$) are made $z$ dependent. 

\subsection{Shaping of stationary light pulses using the optical comb 
technique}

Let us first consider the case when the atomic sample is
irradiated with several counter-propagating 
control fields, with detunings $\Delta_k$, 
and (complex) Rabi frequencies $\Omega_{\pm k}$, such that
the slowly-varying total Rabi frequency reads 
$\Omega(z,t)=\sum_{k}\left(\Omega_{+ k} {\rm e}^{-i\Delta_k(t+z/c)}
+\Omega_{- k} {\rm e}^{-i\Delta_k(t-z/c)}\right)$. 
$\Delta_k=\omega_k-\omega_{ab}$ is the detuning from the atomic resonance.
For simplicity we here consider a degenerate level scheme, i.e.
$\omega_{ab}=\omega_{ac}$.
We assume equal intensities of the corresponding forward and backward
components $|\Omega_{+k}|=|\Omega_{-k}|$.
Corresponding to these driving fields are signal field slowly varying 
amplitudes ${E}_{\pm k}(z,t)$, so that the total signal field envelope is 
$E(z,t)=\sum_k\left({E}_{+k}(z,t){\rm e}^{-i\Delta_k(t-z/c)}
+{E}_{-k}(z,t){\rm e}^{-i\Delta_k(t+z/c)}\right)$.

Assuming $|\Delta_{k\ne 0}|\gg|\Omega_{k\ne 0}|$ so 
that we can ignore coupling of the 
frequency components through off-resonant processes, we may also expand 
the optical polarization as 
$\sigma_{ba}(z,t)=\sum_k\left(\sigma_{ba}^{(+k)}(z,t)
{\rm e}^{-i\Delta_k(t-z/c)}+\sigma_{ba}^{(-k)}(z,t)
{\rm e}^{-i\Delta_k(t+z/c)}\right)$. 
Within the weak-probe and secular approximation 
and assuming two-photon resonance of all
associated pairs of fields we obtain the following equations of motion
\begin{eqnarray}
\dot\sigma_{ba}^{(\pm k)} &=& - \bigl(i\Delta_k+\gamma\bigr)
\sigma_{ba}^{(\pm k)}
+ i g N E_{\pm k} + i\Omega_{\pm k}
\sigma_{bc},\label{Langevin-21}\\
\dot\sigma_{bc} &=& 
+ \sum_k \left[i\Omega_{+k}^*\sigma_{ba}^{(+k)}
+ i\Omega_{-k}^*\sigma_{ba}^{(-k)}\right].\label{Langevin-22}
\end{eqnarray}
Here again the Langevin noise operators associated with the
spontaneous decay from the excited state in eqs.(\ref{Langevin-21})
have been neglected as they do not contribute in the adiabatic
limit. The atomic polarizations $\sigma_{ba}^{(\pm k)}$ drive the
probe field components $E_{\pm k}$ through the shortened one-dimensional
wave equations
\begin{equation}
\Bigl(\partial_t\pm c\partial_z\Bigr)\, E_{\pm k}(z,t) = i g \sigma_{ba}^{(\pm k)}.\label{Maxwell-comb}
\end{equation}

Solving eq.(\ref{Langevin-21}) 
adiabatically for $\sigma_{ba}^{(\pm k)}$ yields
\begin{equation}
\sigma_{ba}^{(\pm k)}=\frac{igN}{\Gamma_k} E_{\pm k}
+\frac{i\Omega_{\pm k}}{\Gamma_k} \sigma_{bc},
\end{equation}
where we have introduced the notation $\Gamma_k=\gamma+i\Delta_k$.
Substituting this result into eq.(\ref{Langevin-22}) for the
ground-state coherence leads to
\begin{eqnarray}
\dot\sigma_{bc} = -\sum_k\frac{|\Omega_{+k}|^2+|\Omega_{-k}|^2}{\Gamma_k}
\sigma_{bc} -g N\sum_k\frac{\Omega_{+k}^*E_{+k}+\Omega_{-k}^*E_{-k}}{\Gamma_k}.
\end{eqnarray}
Letting $\Delta_0=0$ and $|\Delta_{k\ne 0}| \gg \gamma, |\Omega_{\pm k}|$, 
we find that the spin coherence is driven only by the resonant fields, i.e.
\begin{equation}
\dot\sigma_{bc} \approx -  \frac{\Omega_{0}^2}{\gamma} \sigma_{bc}
-gN\frac{\Omega_{+0}^*E_{+0}+\Omega_{-0}^*E_{-0}}{\gamma},
\end{equation}
where $\Omega_0^2=|\Omega_{+0}|^2+|\Omega_{-0}|^2$. In a similar way as in
Sec.2.2 we can solve this equation in first order of an adiabatic expansion.
This yields 
\begin{equation}
\sigma_{bc} = -g N\frac{E_{+0} \Omega_{+0}^* +E_{-0} \Omega_{-0}^*}{\Omega_0^2}
 +\frac{\gamma}{\Omega_0^4} gN \Bigl(\Omega_{+0}^*\partial_t E_{+0}
+\Omega_{-0}\partial_t E_{-0}\Bigr).
\end{equation}
Substituting this result first 
into the equations for the {\sl resonant} fields $E_{\pm 0}$
leads to similar equations as in Sec.2.
\begin{eqnarray}
&&\Bigl(\partial_t \pm c\cos^2\theta\partial_z\Bigr)E_{\pm 0}
= \pm\frac{\sin^2\theta}{\Omega_0^2}\Omega_{\mp 0}^*
\bigl(\Omega_{-0}\partial_t E_{+0} -
\Omega_{+0}\partial_t E_{-0}
\bigr) \label{Maxwell-E0}\\
&&\qquad\qquad
\mp \cos^2\theta\frac{g^2 N}{\Omega_0^2 \gamma}\Omega_{\mp 0}^* 
\bigl(\Omega_{-0}E_{+0}- \Omega_{+0}E_{-0}\bigr).\nonumber
\end{eqnarray}
Taking into account that $|\Omega_{-k}|=|\Omega_{+k}|$ these equations can be 
written in a simpler form introducing sum and difference normal modes
$E_S=(\Omega_{+0}^*E_{+0}+\Omega_{-0}^*E_{-0})/\Omega_0,$ 
 $E_D=(\Omega_{-0}E_{+0}-\Omega_{+0}E_{-0})/\Omega_0$:
\begin{eqnarray}
\partial_t E_S &=& - v_{\rm gr} \partial_z E_D,\\
\partial_t E_D &=& - c \partial_z E_S - \frac{g^2 N}{\gamma} E_D.\label{eq:ed}
\end{eqnarray}
The second term on the right hand side of eq.(\ref{eq:ed}) leads to a
fast decay of the difference mode, such that in the long-time limit
the resonant probe-field amplitudes are matched to the corresponding
drive-field amplitudes 
\begin{equation}
\Omega_{-0}E_{+0}=\Omega_{+0}E_{-0} = - \frac{\Omega_{-0}\Omega_{+0}}{g N}
\sigma_{bc}.\label{resonant}
\end{equation}
As in section 2, the sum normal mode undergoes a diffusion process
under conditions when the difference normal mode can be adiabatically
eliminated. As a consequence the spin excitation, which adiabatically
follows the sum normal mode, does the same, i.e.
\begin{equation}
\partial_t \sigma_{bc}(z,t) = D  \partial_z^2 \sigma_{bc}(z,t),
\label{spin-diffusion}
\end{equation}
with $D=v_{\rm gr} l_{\rm abs}$. 

For the {\sl non-resonant} probe-field components $E_{\pm k}$, $k\ne 0$
one finds the shortened wave-equations 
\begin{equation}
\Bigl(\partial_t\pm c\partial_z\Bigr)\, E_{\pm k}(z,t) = - \frac{g^2 N}{\Gamma_k} E_{\pm k} - \frac{g \Omega_{\pm k}}{\Gamma_k}\sigma_{bc}.
\end{equation}
The second term on the right hand side does not depend on $E_{\pm k}$.
As a consequence, assuming a sufficiently dense medium, the off-resonant
probe amplitude may be adiabatically eliminated leading to
\begin{equation}
E_{\pm k} = - \frac{\Omega_{\pm k}}{g N}\sigma_{bc}.
\end{equation}
Noting that a similar relation holds for the resonant components in the
long-time limit, eq.(\ref{resonant}),  we finally arrive at
\begin{eqnarray}
E(z,t) &=&\sum_k\left(E_{+k}(z,t) {\rm e}^{-i\Delta_k(t-z/c)}
+E_{-k}(z,t) {\rm e}^{-i\Delta_k(t+z/c)}\right)\nonumber\\
&=&  -\sum_k\left(\Omega_{+k} {\rm e}^{-i\Delta_k(t-z/c)}
+\Omega_{-k} {\rm e}^{-i\Delta_k(t+z/c)}\right) \frac{\sigma_{bc}(z,t)}{g N},\\
&=& - \Omega(z,t)\frac{\sigma_{bc}(z,t)}{g N}.\nonumber
\end{eqnarray}
Thus the electric field envelope $E(z,t)$ becomes matched to the control field 
envelope $\Omega(z,t)$ 
modified by the spin coherence $\sigma_{bc}(z,t)$. This allows to control the 
stationary pulse shape through control of 
the retrieve amplitudes and phases. 

Note that the condition for this analysis to hold is $l_{\rm spin}/l_{\rm abs}
\gg$ max$\{\Delta_k\}/\gamma$. Taking the length of 
the initial spin excitation $l_{\rm spin}$ 
to be such that the pulse just fits inside the medium $l_{\rm spin}\sim l$, 
this condition implies that the maximum frequency detuning $\Delta_{max}$ 
for which the stationary pulse can adiabatically follow the control field 
through pulse matching, is given by $\Delta_{max}\sim\gamma \, OD$, 
where $OD=\frac{g^2N l}{\gamma c} =l/l_{\rm abs}$ is the on-resonance 
optical depth. Thus spatial features as small as the optical depth 
can be imposed on the stationary pulse through the frequency comb 
technique. 

It should be noted, however, that the generation of spatially narrow
stationary fields by means of 
the frequency-comb technique is a filtering process
rather than a compression of excitation. In fact the total number of
probe photons created by a frequency comb is much less than in the case
when only the resonant components $\Omega_{\pm 0}$ of the retrieve laser
are present. The excitation density at the center of the stationary
photon wavepacket is the same in both cases, while in the wings it is
substantially smaller for the case of the frequency comb as compared to the
case of homogeneous retrieve beams.

\subsection{Shaping of stationary light pulses using a spatially varying group
velocity}

Let us now discuss the second method indicated in fig.~\ref{fig:cavity} 
in detail.
Assuming again single and two-photon resonance and an optically thick
medium, we can adiabatically eliminate the difference normal mode from
(\ref{Maxwell-E-D}). This yields
\begin{equation}
E_D\approx -\sin(2\phi) l_{\rm abs}\partial_z E_S -  \cos(2\phi) 
l_{\rm abs} \left(\partial_z \phi\right)\, E_S,
\end{equation}
where we have assumed that $\phi$ changes only little over the absorption
length $l_{\rm abs}$ and thus $|(g^2 N/\gamma)|\gg |(\partial_z\phi)
\sin(2\phi)|$. Substituting this into (\ref{Maxwell-E-S}) gives
\begin{eqnarray}
\frac{\partial}{\partial t} E_S 
= A_0 \, E_S + \frac{\partial}{\partial z}\Bigl[A_1\, E_S\Bigr]
+\frac{\partial^2}{\partial z^2}\Bigl[D\, E_S\Bigr],
\label{almost-FPE}
\end{eqnarray}
where $D=v_{\rm gr} l_{\rm abs}$ is the diffusion 
constant introduced before, and the coefficients $A_0$ and $A_1$ read
\begin{eqnarray}
A_0&=& -v_{\rm gr}\bigl[(\partial_z\phi)\sin(2\phi)
+2 l_{\rm abs}(\partial_z\phi)^2\sin^2(2\phi)\nonumber\\
&&\qquad -l_{\rm abs}
(\partial_z\phi)^2\cos^2(2\phi)-(\partial^2_z\phi)\cos(2\phi)\sin(2\phi)\bigr],
\\
A_1&=&-v_{\rm gr}\bigl[1+4l_{\rm abs} (\partial_z\phi)\bigr].
\end{eqnarray}
The constant term proportional to $A_0$ in eq.(\ref{almost-FPE}) 
can be removed by the substitution
\begin{equation}
E_S=\widetilde E_S\, \exp\bigl\{A_0 t
\bigr\},\label{tilde-E-S}
\end{equation}
which results into a  Fokker-Planck equation for $\widetilde E_S$:
\begin{equation}
\frac{\partial}{\partial t} \widetilde E_S =
\frac{\partial}{\partial z}\Bigl[A_1\, \widetilde E_S\Bigr]
+\frac{\partial^2}{\partial z^2}\Bigl[D\, \widetilde E_S\Bigr].
\label{FPE}
\end{equation}
In the following we will discuss the spatio-temporal evolution of $E_S$ 
resulting from this equation.

Non equal drive fields lead to an effective group 
velocity $v_{\rm gr}\cos(2\phi)$ for the sum normal mode. 
If this group velocity is tailored in such
a way that it is negative for positive values of $z$ and positive for
negative values of $z$, there is an effective drift
towards the origin. This force may compensate the dispersion due to the
absorption of large-$k$ components of the probe field found in section 3.
We thus consider as the simplest example  
the special case of a linearly varying intensity
difference of the two drive fields with a constant sum $\Omega_0^2=$ const.: 
\begin{equation}
\cos(2\phi)=\frac{|\Omega_+|^2-|\Omega_-|^2}{\Omega_0^2} 
\approx -\frac{z}{l},\qquad \sin(2\phi) \approx 1\, .
\label{linear}
\end{equation}
This situation is realized e.g. if the
two control fields are paraxial, Gaussian laser beams with focal points at 
$z_{\pm}\approx \pm 2 l$. 
%
%
\begin{figure}[hbt]
\begin{center}
\includegraphics[width=7.5cm]{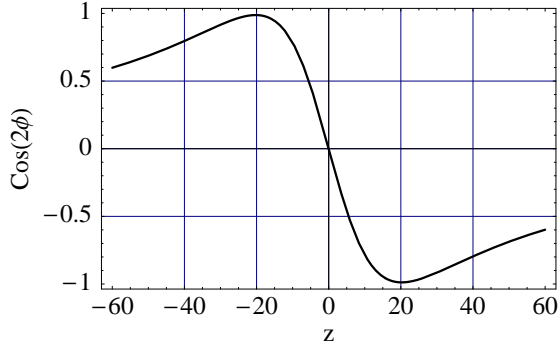}
\caption{The prefactor of the drift term for the configuration 
of to separated Gaussian beams. 
The focal points of the two beams are located at 
$z_{\pm}=\pm 20\, l_{\rm abs}$. The linear regime 
under consideration in this 
section is clearly visible.
\label{LinearField}
}
\end{center}
\end{figure}
%
%
The linear approximation is of course only valid
for $|z|\ll l$.
In this case eq.(\ref{FPE}) 
turns into the
Fokker-Planck equation of the Ornstein-Uhlenbeck process
\cite{Gardiner-Handbook} for which exact analytic solutions are 
known
\begin{equation}
\frac{\partial}{\partial t} \widetilde E_S =
\frac{\partial}{\partial z}\Bigl[\, v_{\rm gr} \frac{z}{l}\widetilde E_S\Bigr]
+\frac{\partial^2}{\partial z^2}\Bigl[D\, \widetilde E_S\Bigr].
\label{Ornstein-Uhlenbeck}
\end{equation}
Here we have neglected contributions proportional to $l_{\rm abs}/l$ as
compared to unity.
The Ornstein-Uhlenbeck process has a stationary Gaussian solution with width
$\sqrt{l\, l_{\rm abs}}$. 
Noting that now $A_0=- v_{\rm gr}(\partial_z\phi)=
-v_{\rm gr}/2l$ this gives with eq.(\ref{tilde-E-S}) in 
the long-time limit:
\begin{equation}
E_S(z,t) =
\widetilde E_S(z)\, \exp\bigl\{A_0 t\bigr\}
\longrightarrow
\exp\left\{-\frac{z^2}{{2} l l_{\rm abs}}\right\} 
\exp\left\{-\frac{v_{\rm gr} t}{2 l}\right\}.
\label{StationarySolutionOUProcess}
\end{equation}
The use of retrieve lasers with non-equal and spatially varying intensities
thus acts like an effective cavity for the probe field with a ring-down
time $l/v_{\rm gr}$ 
given by the time a photon travels between the intensity maxima of the 
two drive lasers. 

The initial-value problem of the Ornstein-Uhlenbeck process can be solved
by making use of the eigensolutions $\{\Phi_n(z),\lambda_n\}$ of the 
corresponding
backward equation \cite{Gardiner-Handbook}
\begin{equation}
\frac{\partial^2}{\partial z^2}\, \Phi_n(z)-\frac{v_{\rm gr}}{D} \frac{z}{l}\frac{\partial}{\partial z}\, \Phi_n(z) 
  +\frac{\lambda_n}{D}\, \Phi_n(z)=0.
\label{Ornstein-Uhlenbeck-EV}
\end{equation}
%
%
%
%
Eq.(\ref{Ornstein-Uhlenbeck-EV}) is
the differential equation of the Hermite polynomials $H_n$ and thus the
eigenvalues $\lambda_n$ and eigenfunctions $\Phi_n(z)$ read
\begin{eqnarray}
\lambda_n &=& n\, \frac{v_{\rm gr}}{l},\qquad\qquad n\in \{0,1,2,\dots\},\\
\Phi_n(z) &=& (2^n n!)^{-1/2}\, 
H_n\Bigl(\frac{z}{\sqrt{2 l\, l_{\rm abs}}}\Bigr).
\end{eqnarray}
The general solution of the initial value problem
then reads
\begin{eqnarray}
&&E_S(z,t) =\sum_{n=0}^\infty
 \frac{c_n}{\sqrt{2^{n+1}n!\pi\, l\, l_{\rm abs}}}\, 
\exp\left\{-\frac{z^2}{2\, l\, l_{\rm abs}}\right\} \times \nonumber\\
&&\qquad\qquad\quad\times \,H_n\Bigl(
\frac{z}{\sqrt{2\, l\, l_{\rm abs}}}\Bigr)\,
\exp\left\{-\frac{v_{\rm gr}(n+1/2) t}{l}\right\}.
\label{initial-value-solution}
\end{eqnarray}
The coefficients $c_n$ are determined by the initial 
field $E_S(z,0)$:
\begin{equation}
c_n = \int_{-\infty}^\infty\!\! {\rm d}z\, E_S(z,0)\, 
 H_n\Bigl(\frac{z}{\sqrt{2\, l\, l_{\rm abs}}}\Bigr)\, (2^n n!)^{-1/2}.
\label{coefficients}
\end{equation}
It is interesting to note that, apart from the additional overall damping term,
eq.(\ref{initial-value-solution}) is  very
similar to a damped harmonic oscillator with oscillator length  
$\sqrt{l\, l_{\rm abs}}$.
If the stored light pulse is a Gaussian and if the separation of the 
foci of the two retrieve lasers is chosen such that the width of the 
stored light pulse is less than the effective oscillator length $
\sqrt{ l\, l_{\rm abs}}$, only the
fundamental mode $\Phi_0$ gets excited in the retrieve process.
In this case a spatially constant field distribution is created.
The field has however a finite lifetime determined by the decay rate
$\gamma_{\rm eff}=v_{\rm gr}/ l$. 
As a consequence the total excitation decays in time according to
\begin{equation}
n_{\rm tot}(t)=n_{\rm tot}(0)\, \exp\bigl\{- v_{\rm gr} t/ l\bigr\}.
\end{equation}
In order to have negligible losses, the time over which the stationary
pulse can be maintained is limited by exactly the same expression 
as in the diffusion case
\begin{equation}
t\ll \frac{l}{v_{\rm gr}} = \frac{\Delta z^2(0)}{v_{\rm gr} l_{\rm abs}}.
\end{equation}

If the separation of the focal points in the retrieve process is much smaller 
than $l_{\rm spin}^2/l_{\rm abs}$ the generated stationary light pulse
has a much narrower width than the original spin excitation 
($l_{\rm spin}$). In this 
case a large number of higher-order Gauss-Hermite modes is excited
however (see eq.(\ref{coefficients})), which decay much faster than the
fundamental mode. Thus as in the case of the frequency comb very narrow
spatial distributions of the stationary field can be created, however
only through a filter process.

\begin{figure}[htb]
\begin{center}
\includegraphics[width=12.0cm]{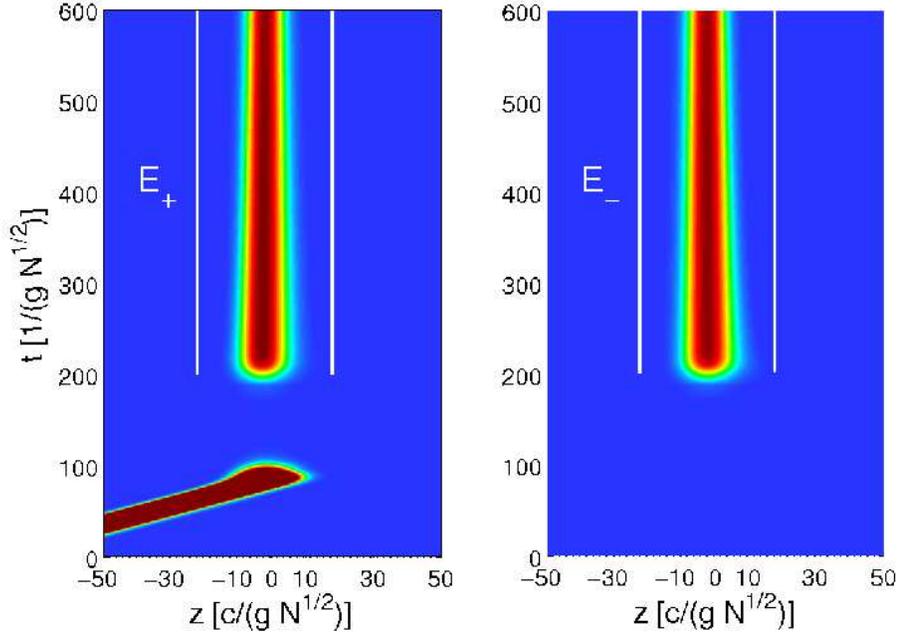}
\caption{Storage of a Gaussian pulse and subsequent retrieval with
two control beams with spatially changing intensity difference. 
The maxima of the
intensities of $\Omega_\pm$ are indicated by the two white lines.
Close to the midpoint between these lines the normalized intensity
difference $(|\Omega_{+0}|^2- |\Omega_{-0}|^2)/\Omega_0^2$
varies linearly with $z$. The generation of fields with constant spatial
shape is apparent. The parameters are that of fig.(\ref{fig-2}). 
In addition Gaussian beam profiles 
$\Omega_\pm(z)=1/w_\pm (z)$ with $w_\pm (z)=\sqrt{1-2(z-z_\pm)/\pi}$
have been assumed.
\label{fig-4}}
\end{center}
\end{figure}

In fig.\ref{fig-4} a numerical simulation of the 
the retrieval using two control beams with separated foci is shown. 
Initially a Gaussian probe pulse is stored in a collective spin excitation.
The center of the stored probe pulse is in the middle between
the two foci, indicated by the two white lines and the width of the 
pulse is on the order of $\sqrt{l\, l_{\rm abs}}$. Thus mainly the
fundamental mode $\Phi_0$ is excited, which can be seen from the
constant spatial shape of the retrieved wave-packet.


\section{Spatial compression of stationary pulses of light}

\begin{figure}[htb]
\begin{center}
\includegraphics[width=12.0cm]{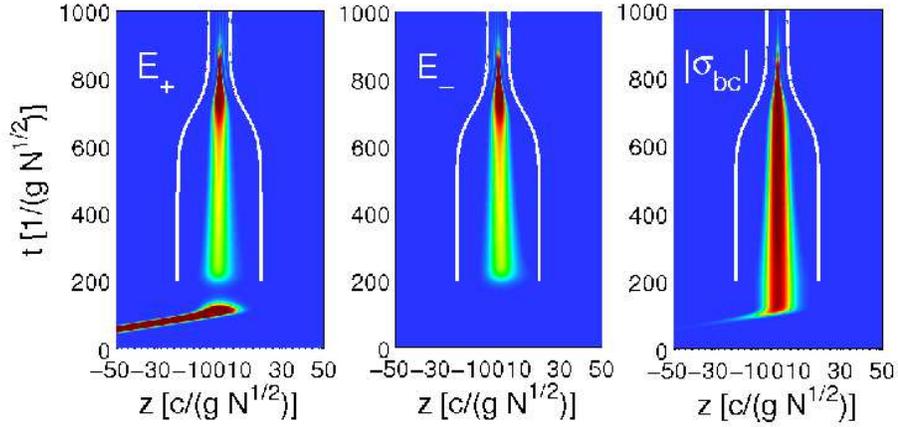}
\caption{Retrieval of a stored pulse and spin excitation 
using drive fields with separated
foci and subsequent reduction of their separation. The position of the
focal points is indicated by white lines. One clearly recognizes
a compression of the probe field associated with an increase of the
field density. The parameters are the same as in fig.(\ref{fig-4}) 
except for $\gamma=0.05$. 
The foci move like: $z_\pm(t)= 20\mp 10*0.5*[1+\tanh(0.0125*(t-700))]$}.
\label{fig-5}
\end{center}
\end{figure}

We have seen in the last section that the use of retrieve lasers with 
spatially modulated intensities does allow the generation of stationary
light pulses with very narrow spatial shapes. The underlying process is
however a {\it filter process} and thus accompanied 
either by reduction of the photonic component in the polariton or
by large losses. 
Nevertheless both techniques open interesting possibilities
for the {\it spatial compression} of a stored photon with small losses.
If a stationary light pulse is created e.g. with separated foci of the 
drive lasers as explained
in the previous section, and the distance between the focal points is
reduced as a function of time in an adiabatic way, the spatial width 
of the stationary pulse follows. This results in a spatial compression
of the probe excitation. As the effective decay rate of a stationary
pulse $\gamma_{\rm eff}$ increases with decreasing pulse width, the control
fields should be switched off immediately after the compression. 
Fig. \ref{fig-5} shows a numerical example for such a process.
After retrieval of a stored pulse with separated foci, whose position
is indicated by white lines, the distance between the focal points is
reduced. One can see very clearly that the field density as well as the
density of the spin excitation is substantially 
increased in this process. 
This suggests that by changing the spatial profile of the control field in time, either using the optical comb technique or displaced foci, spatial compression can be achieved.

\begin{figure}[htb]
\begin{center}
\includegraphics[width=9.0cm]{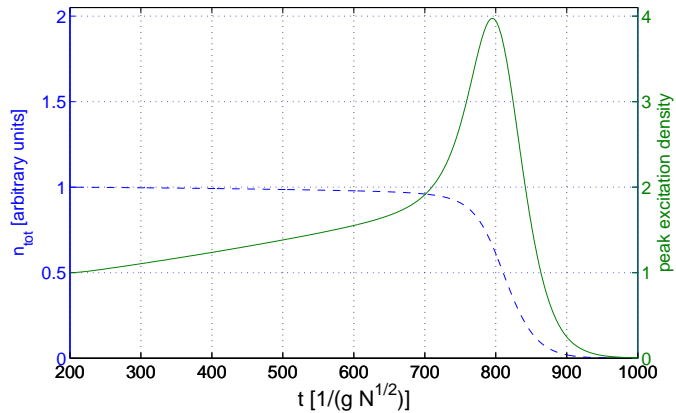}
\caption{Peak excitation density (solid line) and total excitation (dashed 
line) as function 
of time for the example of fig.\ref{fig-5}. All quantities are 
normalized to their value at $t=200 (g\sqrt{N})^{-1}$.
}
\label{fig-6}
\end{center}
\end{figure}

In fig.\ref{fig-6} the temporal evolution of the peak 
density and the total excitation
(i.e. photon number of the stationary field plus spin excitations in
the atomic ensemble) are shown for the example of fig.\ref{fig-5}. One
recognizes that when the peak density increases the total number of 
excitations starts to decrease very rapidly. Clearly an 
optimization is needed to maximize fidelity and compression. 
In order to find optimum conditions and to
estimate the maximum possible compression from the theoretical model,
it is necessary to
include non-adiabatic couplings into the description.
Finally, it is not clear if spatial compression of this kind can be used to enhance further nonlinear optical processes using the techniques of Ref. \cite{Andre-PRL-2005}.
This analysis is beyond the scope of the present paper and will be discussed
in detail elsewhere.

\section{Summary and outlook}

In the present paper we have discussed the generation and coherent control 
of stationary
pulses of light by storage of a light pulse in a collective spin excitation
via EIT and subsequent partial retrieval of this excitation with 
counter-propagating retrieve lasers. We have shown that for equal 
and spatially homogeneous intensities of the
forward and backward retrieve beams a quasi-stationary wavepacket of the
probe light is generated with an initial envelope given by the spin excitation.
For an optically thick medium the dynamics of the wavepacket is described
by a diffusion equation with a diffusion constant given by the product
of group velocity and absorption length without EIT. The physical mechanisms
for the diffusion of the stationary pulse is the well-known phenomenon
of pulse matching of probe and retrieve field components in EIT. 
We have shown furthermore that spatially modulated retrieve lasers 
can be used to manipulate the shape of the stationary light pulse 
and in particular to spatially compress the excitation. Making use of
a frequency comb for the retrieve fields a very narrow spatial distribution
of the probe field can be generated. Likewise the use of retrieve fields
with spatially varying intensity difference can lead to a narrow non-dispersing
but exponentially decaying field distribution. In both cases the 
narrow field distribution is however created by a filtering process. 
Thus these techniques can not straightforwardly be
applied, e.g. to applications in quantum nonlinear optics.
However, as demonstrated with
some numerical examples, if the retrieve 
field distribution is initially matched to the stored spin excitation and 
its shape is modified in time in an adiabatic way, the stationary light pulse
and thus the stored excitation can be spatially compressed. Although we 
have not analyzed the fidelity of the compression process quantitatively and
have not optimized it, the present paper shows the potentials of
coherent control of stationary light pulses for quantum nonlinear optics and quantum information processing
with photons and atomic ensembles.

\section*{Acknowledgement}

M.F. and F.Z. would like to thank the Institute for Atomic Molecular
and Optical Physics at the Harvard-Smithsonian Center for Astrophysics
as well as the Harvard physics department for their hospitality and
their support. The financial support of the Graduiertenkolleg 792 at the
Technical University of Kaiserslautern is gratefully acknowledged.
The work at Harvard is supported by NSF, DARPA, the Alfred P. Sloan Foundation, and the David and Lucile Packard Foundation.


\bibliographystyle{unsrt}


\appendix
\section{Multi-component Spatial Coherence Grating}

In this appendix, we analyze the stationary pulse solutions from the point of view of spatial coherence gratings \cite{moiseev05a,moiseev05b}, and we show that multiple atomic momentum components can be taken into account. These arise due to multiple scattering of photons in the forward and backward directions, resulting in distinctly different atomic susceptibilities. For stationary atoms, such as in cold atomic samples, these multiple scattering momentum components can be populated and preserve their coherence. In contrast, for warm atomic vapors, the rapid random motion of the atoms and their collisions results in a very rapid decay of spatial coherences with period equal to or shorter than the optical wavelength. 

As discussed in section 2.1,
associated with the forward/backward
propagating fields are slowly varying optical coherences (polarization) with slowly varying 
envelopes $\hat{P}_\pm(z,t)$, so that the total polarization can be written as
$\sqrt{N}\hat{\sigma}_{ba}(z,t)=\hat{P}_+(z,t)e^{ik_cz}+\hat{P}_-(z,t)
e^{-ik_cz}$.
Similarly,
the ground-state spin coherence is defined as $\mathcal{S}(z,t)=\sqrt{N}\sigma_{bc}(z,t)$.
Letting the wave-vector mismatch be $\Delta K=k_c-k$, and defining $\mc{E}_\pm=E_{\pm}e^{\mp i\Delta Kz}$ and $\mc{P}_\pm=P_\pm e^{\mp i\Delta Kz}$, the equations of motion for the fields can then be written as
\begin{subequations}
\label{eq:splcnteqns}
\begin{align}
(\partial_t+c\partial_z)\mc{E}_+(z,t) &= i\Delta Kc\,\mc{E}_++ig\sqrt{N}\mc{P}_+&
\\
(\partial_t-c\partial_z)\mc{E}_-(z,t) &= i\Delta Kc\,\mc{E}_-+ig\sqrt{N}\mc{P}_-&
\end{align}
\end{subequations}
while the atomic equations of motion are (setting $\Delta=\delta=0$)
\begin{subequations}
\label{eq:splateqns}
\begin{align}
\partial_t\mc{P}_+ &= -\gamma\mc{P}_+
+i\Omega_{+0}\mc{S}+ig\sqrt{N}\mc{E}_+ &
\\
\partial_t\mc{P}_- &= -\gamma\mc{P}_-
+i\Omega_{-0}\mc{S}+ig\sqrt{N}\hmc{E}_- &
\\
\partial_t\hmc{S} &= i\Omega_{+0}^*\mc{P}_+
+i\Omega_{-0}^*\mc{P}_-. &
\end{align}
\end{subequations}

These equations show that the counter-propagating control fields $\Omega_{\pm0}$ induce a coupling between $\mc{E}_+$ and $\mc{E}_-$, mediated through the spin coherence $\mc{S}$.
This leads to the formation of new eigenmodes of propagation, where as we show below, there is one mode that is very rapidly decaying while the other decays very weakly in the large optical depth limit. 
This phenomenon is analogous to the ``pulse matching'' phenomenon, as first described by Harris \cite{harris93}. 
These equations can be used to obtain the susceptibilities, given by $\frac{\omega_{ab}}{2}\chi_{\sigma\sigma'}=g\sqrt{N}\mc{P}_{\sigma}/\mc{E}_{\sigma'}$, where $\sigma,\sigma'=\pm$.

We now contrast two approaches to compute the susceptibilities, one approach in which the secular approximation is made and one where it is not. 
Writing the polarization as $\mathcal{P}(z,t)=
\sum_{n}\mathcal{P}_{2n+1}(z,t)e^{i(2n+1)k_cz}$, and the spin wave be 
$\mathcal{S}(z,t)=\sum_{n}S_{2n}(z,t)e^{2ink_cz}$, we have
the effective Hamiltonian
\ba
H &=& \frac{-1}{L}\int dz\left[g\sqrt{N}
(\mathcal{E}_+\mathcal{P}_{1}^\dagger+\mathcal{E}_-\mathcal{P}_{-1}^\dagger)
\right.\nonumber \\
&+& \left. \sum_{n}(\Omega_{+0}\mathcal{S}_{2n}\mathcal{P}_{2n+1}^\dagger
+\Omega_{-0}\mathcal{S}_{2n}\mathcal{P}_{2n-1}^\dagger)+h.c.\right].
\ea

The equations of motion for the fields are given by
\ba
(\partial_t+c\partial_z)\mathcal{E}_+ &=& i\Delta Kc\mathcal{E}_+
+ig\sqrt{N}\mathcal{P}_{1}
\\
(\partial_t-c\partial_z)\mathcal{E}_- &=& i\Delta Kc\mathcal{E}_-
+ig\sqrt{N}\mathcal{P}_{-1},
\ea
while the atomic equations of motion are
\ba
\partial_t\mathcal{P}_{2n+1} &=& -\gamma\mathcal{P}_{2n+1}
\nonumber \\
&+& i\Omega_{+0}\mathcal{S}_{2n}+i\Omega_{-0}\mathcal{S}_{2(n+1)}
+ig\sqrt{N}\left(\delta_{n,0}\mathcal{E}_{+1}+\delta_{n,-1}\mc{E}_{-1}\right)
\label{eq:coupeqn1}
\\
\partial_t\mathcal{S}_{2n} &=& -i(\Omega_{+0}^*\mathcal{P}_{2n+1}
+\Omega_{-0}^*\mathcal{P}_{2n-1})
\label{eq:coupeqn2}
\ea

\begin{figure}[ht]
\begin{center}
\includegraphics[width=115mm]{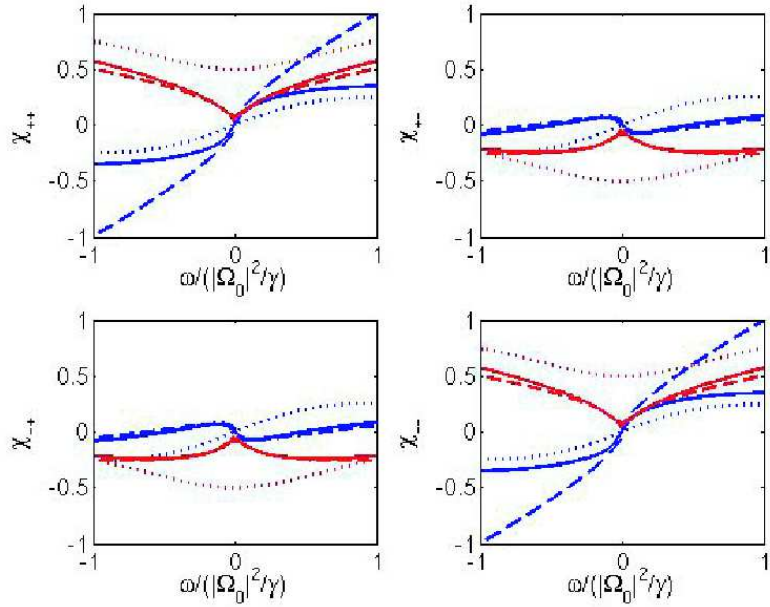}
\end{center}
\caption{\label{fig:multisuscept}
Self- and cross-susceptibilites (arbitrary units) $\chi_{\pm,\pm}(\omega)$ vs. frequency $\omega$ (in units of $|\Omega_0|^2/\gamma=(|\Omega_+|^2+|\Omega_-|^2)/\gamma$), for $\Omega_+=\Omega_-$,
$\Omega_0=0.1\gamma$.
Real part (blue) and imaginary part (red), dotted line: Multiple spatial component approach truncated at $\mc{P}_{\pm 1}$, dashed line: Coupled Mode equations (analytic result), full line: Multiple spatial component approach truncated at $\mc{P}_{\pm(2n+1)}$ with $n=5$.}
\end{figure}

The susceptibilites $\chi_{\sigma\sigma'}(\omega)$ can be computed from these coupled equations. Truncating the decomposition of the polarization at $\mc{P}_{\pm(2n+1)}$, we can easily show that $n=0$ reproduces our earlier result of section 2.1, whereas letting $n\rightarrow\infty$ leads to a different limit  (see Fig.~\ref{fig:multisuscept}). 
We thus find that there is a clear difference between the multi-component case and the case when only the zero momentum component of the spin wave contributes. These two situations correspond e.g. to cold atomic samples vs. hot atomic gases, so that based on the previous considerations we expect that the stationary light pulse will be very different in cold atomic samples. 

The susceptibilities obtained in the limit $n\rightarrow\infty$ can be shown \cite{Andrephd} to be the same as those found from a coupled mode approach \cite{Yariv}.
Maxwell's equation in a spatially modulated medium in 1D is (in the frequency domain)
\ba
c^2\frac{d^2}{dz^2}E(z,\nu)+\nu^2[1+\chi(z,\nu)]E(z,\nu) &=& 0.
\ea
Putting $E(z,\nu)=E_+(z,\omega)e^{ik_cz}+E_-(z,\omega)e^{-ik_cz}$, 
where $\omega=\nu-\omega_c$,
coupled mode equations can be obtained by letting the susceptibility be given by the usual EIT susceptibility \cite{Lukin-RMP-2003}
\ba
\chi(z,\omega) &=& \frac{2g^2N}{\gamma\omega_0}
\frac{i\gamma\Gamma_0}{\Gamma\Gamma_0+|\Omega(z)|^2}
\label{eq:chicold}
\ea
where the control field amplitude is space dependent $\Omega(z)=\Omega_{+0}e^{ik_cz}+\Omega_{-0}e^{-ik_cz}$, and where
$\Gamma=\gamma-i\omega$, and $\Gamma_0=\gamma_0-i\omega$ ($\gamma_0$ is the ground state coherence decay rate, which we ignore for simplicity  in this work).
Using the Fourier expansion 
$\chi(z,\omega)=\sum_{n}\chi_n(\omega)e^{ink_cz}$ and using coupled mode 
analysis \cite{Yariv}, we can compute the susceptibilities $\chi_{\sigma,\sigma'}(\omega)$. The result of this calculation is compared 
to the result of the multiple spatial component approach outlined 
previously, in Fig.~\ref{fig:multisuscept}.

Implicit in the derivation of the susceptibilities with a coupled mode approach, is the assumption of stationary atoms, so that all intermediate atomic momentum states are taken into account, in particular high spatial frequency  coherences contribute to this result.
All spatial components of the spin and polarization waves are taken into 
account, even those with momentum equal to multiples of the optical wavevector $K$.
For warm atomic samples, these coherences rapidly decohere due to the random motion of the atoms and their collisions. Hence, only for cold atoms, such as those in a Bose-Einstein condensate, these spatial coherences may play a role and lead to interesting differences with stationary light phenomenon observed in thermal atomic vapors.

Finally, we argue that for storage and retrieval of excitations, these large spatial wavevector coherences are mostly irrelevant. As can be seen from \ref{eq:coupeqn1} and \ref{eq:coupeqn2}, the signal field (in modes $\mc{E}_{\pm}$) couples to the polarization components $\mc{P}_{\pm 1}$ with Rabi frequency $g\sqrt{N}$, which in the ``slow'' light limit is much larger than the control field Rabi frequencies $\Omega_{\pm0}$. 
Therefore, starting from a stored pulse in the zero-momentum spin coherence $\mc{S}_0$, 
most of the amplitude is coupled to the signal field modes $\mc{E}_\pm$, while very little amplitude ``leaks'' to the higher momentum coherences. Thus 
coupling to higher momentum coherences does not lead to decay 
of the spin coherence, as discussed in \cite{Andrephd}.

\end{document}